\documentclass[jcp,aip,showpacs,amsmath,amssymb,unsortedaddress,12p]{revtex4-1}

\usepackage{natbib}
\usepackage[pdftex]{graphicx}
\usepackage{amsfonts}
\usepackage{amssymb}
\usepackage{amsmath}

\usepackage[final]{pdfpages}

\begin{document}

\title{Phase diagram of two-dimensional hard ellipses}

\author{Gustavo Bautista-Carbajal}
\affiliation{Departamento de F\'isica, Universidad Aut\'onoma Metropolitana-Iztapalapa, 09340, M\'exico, Distrito Federal,
Mexico,}
\affiliation{Academia de Matem\'aticas, Universidad Aut\'onoma de la Ciudad de M\'exico, 07160, M\'exico, Distrito Federal,
Mexico. }


\author{Gerardo Odriozola}
\email{godriozo@imp.mx} 
\affiliation{Programa de Ingenier\'{\i}a Molecular, Instituto Mexicano del Petr\'{o}leo, Eje Central
L\'{a}zaro C\'ardenas 152, 07730, M\'{e}xico, Distrito Federal,
Mexico.}

\date{\today}

\begin{abstract}
We report the phase diagram of two-dimensional hard ellipses as obtained from 
replica exchange Monte Carlo simulations. The replica exchange is implemented by 
expanding the isobaric ensemble in pressure. The phase diagram shows four 
regions: isotropic, nematic, plastic, and solid (letting aside the hexatic phase 
at the isotropic-plastic two-step transition [PRL 107, 155704 (2011)]). At low 
anisotropies, the isotropic fluid turns into a plastic phase which in turn 
yields a solid for increasing pressure (area fraction). Intermediate 
anisotropies lead to a single first order transition (isotropic-solid). Finally, 
large anisotropies yield an isotropic-nematic transition at low pressures and a 
high-pressure nematic-solid transition. We obtain continuous isotropic-nematic 
transitions. For the transitions involving quasi-long-range positional 
ordering, i.~e.~isotropic-plastic, isotropic-solid, and nematic-solid, we 
observe bimodal probability density functions. This supports first order 
transition scenarios.  
\end{abstract}


\maketitle

\section{Introduction}
Two-dimensional models are frequently employed as idealizations of quasi-2D 
experimental setups, such as strongly confined colloids or colloidal thin 
films~\cite{Loewen09,Tkalec13,Arciniegas14}. In turn, quasi-2D mesophases and 
nanocrystals can be used as basic units for the synthesis of superlattice 
structures~\cite{Zewei10}, multilayer arrangements by means of layer-by-layer 
assembly~\cite{Schmitt93,Decher97}, and for template assisted assembly 
processes~\cite{Rycenga09}. These applications have encouraged several 
experimental and simulation studies on the behavior of 2D-confined nanocrystals 
of different shapes~\cite{Qi12}. For instance, experiments and simulations have 
shown that needles, squares, octapods, and ellipsoidal anisotropic particles 
produce a rich mesophase behavior when confined to a quasi-2D 
plane~\cite{Frenkel-Eppenga,Cuesta90,Donev06,Avendano12,Shah12,Qi12,Qi13}. For 
designing such arrangements, it is important to take into account the 
directional nature of entropic forces acting on anisotropic 
particles~\cite{Damasceno12,van-Anders14}, i.~e.~the effective forces that 
result from a system's statistical tendency to increase its entropy. In 
particular, for hard systems, these are the only forces acting on the particles 
and are responsible for the different type of phase transitions appearing at 
different densities and particle-anisotropies. Directionality makes anisotropic 
particles show, in general, a much richer phase behavior than isotropic ones. 

Probably the most simple 2D-system is the hard disk model. Nonetheless, the 
phase transition this model shows is, to say the least, hard to elucidate. In 
the first place, the solid phase for 2D-systems has quasi-long-range but not 
true-long-range positional order. That is, positional correlations decay to zero 
following a power-law. On the other hand, bond orientational correlations are 
indeed, long-ranged. Hence, a 2D-solid is not a crystal, since true crystals 
preserve both, bond orientational order and positional order for all distances. 
In the second place, in-between the solid and the liquid, Kosterlitz, Thouless, 
Halperin, Nelson, and Young (KTHNY)~\cite{Kosterlitz-Thouless,Halperin-Nelson} 
proposed the existence of a hexatic phase. This phase is 
characterized by quasi-long-range bond orientational correlations, similar to a 
two-dimensional nematic where orientation is also quasi-long-range 
ordered~\cite{Straley71,Frenkel-Eppenga,Bates-Frenkel,Zheng10,Xu13}, but with a 
sixfold rather than twofold anisotropy. In their scenario, the solid melts into 
a hexatic phase, following a dislocation unbinding process, before turning into 
a liquid by means of disclination unbinding, for decreasing pressure. The theory 
predicts the two transitions to be continuous. The KTHNY two-step continuous 
transition and a single first order transition have been the two of several 
scenarios which have larger support~\cite{Jaster99}. Quite recent long-scale 
computer simulations (containing $1024^2$ particles) strongly suggest a 
liquid-hexatic first order transition, followed by a continuous hexatic-solid 
transition~\cite{Bernard11}. 

A possible way to include anisotropy in the 2D-system is to replace disks by 
ellipses. This looks natural since circles (disks) are a particular type of 
ellipses (with both focal points at the same location), or the other way around, 
ellipses are the generalization of circles. Thus, ellipses on a plane can be 
seen as the most simple anisotropic model in 2D. Studies on ellipses are mostly 
focused on the isotropic-nematic transition~\cite{Xu13}. Monte Carlo simulations 
of Vieillard-Baron are the first of this kind~\cite{Vieillard-Baron72} (there is 
a previous determination of the structure factor of a hard ellipse nematic 
phase~\cite{Levesque69}). For an aspect radio $\kappa=6$ three different phases 
are identified: isotropic, nematic, and solid. For a quasi-spherical case, a 
phase ``analogous to the plastic crystal phase'' was 
found~\cite{Vieillard-Baron72}. Here, plastic crystal means the existence of 
positional order and the lack of orientational order. More evidence on the 
isotropic-nematic transition of hard anisotropic particles, 
needles~\cite{Frenkel-Eppenga} and rods~\cite{Bates-Frenkel}, revealed that this 
type of transition is continuous. In addition, Cuesta and 
Frenkel~\cite{Cuesta90} showed that there is no stable nematic phase in hard 
ellipses for $\kappa=2$. A recent work on hard ellipses by Xu 
et.~al.~\cite{Xu13} gives much further details on the isotropic-plastic and 
isotropic-nematic transition of hard ellipses. In this last work, as well as in 
reference~\cite{Foulaadvand13}, transport properties of fluid phases are also 
studied. However, details on the large area fraction region of the phase 
diagram, which includes both, the isotropic-solid and plastic-solid transitions, 
remain elusive. 

The main goal of this paper is to report the whole phase diagram of hard 
ellipses in the anisotropy region $1 \leq \kappa \leq 5$. For this purpose, we 
are implementing replica exchange Monte Carlo simulations (REMC) by performing a 
pressure extension of the isobaric ensemble. This way, area fluctuations are 
accessed on each set pressure providing useful information at the transitions. 
We detect four phases (letting aside the hexatic phase in-between the isotropic 
and plastic phases~\cite{Bernard11}) which are isotropic, plastic, solid, and 
nematic. At low anisotropies $\kappa \lesssim 1.6$, we have obtained a 
low-pressure isotropic-plastic first order transition (a probable two-step 
transition involving a hexatic phase, with a first order isotropic-hexatic 
transition followed by a subtle and continuous hexatic-plastic transition, 
taking into account Bernard and Krauth conclusions~\cite{Bernard11}) and a 
high-pressure plastic-solid transition. Intermediate anisotropies, $1.6 \lesssim 
\kappa \lesssim 2.4$ yield a single isotropic-solid transition (confirming 
Cuesta and Frenkel results~\cite{Cuesta90} for $\kappa=2$). Finally, for $\kappa 
\gtrsim 2.4$, a low-pressure isotropic-nematic continuous transition and a 
high-pressure nematic-solid transition are found. 

\section{Simulation details}

To detect overlaps we are following the 3D analytical approach of Rickayzen~\cite{Rickayzen98} while restricting the particles geometrical centers to move in a 2D-plane and their axis of revolution to rotate inside it. The Rickayzen-Berne-Pechukas (RBP) expression is given by~\cite{Rickayzen98}
\begin{equation}\label{RBP1}
\sigma_{RBP} = \frac{\sigma_a}{\sqrt{ 1 - \frac{1}{2} \chi \big [ A^{+} + A^{-} \big ] + \big ( 1 - \chi ) \chi' \big [ A^{+} A^{-} \big ]^{\gamma} }},
\end{equation}
where
\begin{equation}\label{RBP2}
A^{\pm} = \frac{ ( \hat{\mathbf{r}} \cdot \hat{\mathbf{u}}_{i} \pm \hat{\mathbf{r}} \cdot \hat{\mathbf{u}}_{j} )^{2} }{ 1 \pm \chi \hat{\mathbf{u}}_{i} \cdot \hat{\mathbf{u}}_{j} },
\end{equation}
\begin{equation}\label{RP2}
\chi = \frac{ \sigma_b^{2} - \sigma_a^{2} }{ \sigma_b^{2} + \sigma_a^{2} } \ \ , \ \ \chi' = \bigg ( \frac{ \sigma_b - \sigma_a }{ \sigma_b + \sigma_a } \bigg )^{2}.
\end{equation}
In this expression, $\sigma_a$ and $\sigma_b$ are the major and minor axes of the ellipses, 
respectively. We take $\sigma_b$ as the length unit and vary $\sigma_a$ 
to obtain different anisotropies. The aspect ratio is given by 
$\kappa=\sigma_a/\sigma_b$, such that $\kappa \geq 1$. $\kappa = 1$ 
corresponds to the disks case. $\hat{\mathbf{u}}_{i}$ and $\hat{\mathbf{u}}_{j}$ 
are unit vectors along the smallest diameters of ellipsoids $i$ and $j$, 
respectively. $\hat{\mathbf{r}}$ is the unit vector along the line joining the 
geometric particle centers. In addition, $\gamma$ is introduced to further 
approach the exact Perram and Wertheim numerical 
solution~\cite{Perram84,Perram85}. $\gamma$ values are given in 
reference~\cite{GuevaraHE}. The average difference between the analytical 
approach and the exact numerical solution is always small~\cite{GuevaraHE}. 

To avoid the inherent hysteresis associated to transitions as far as 
possible~\cite{Bautista13}, we are implementing the replica exchange Monte Carlo 
technique~\cite{Marinari92,Lyubartsev92,hukushima96}. It is based on the 
definition of an extended ensemble whose partition function is given by 
$Q_{ext}=\prod_{i=1}^{n_r}Q_{i}$, being $Q_i$ the partition function of ensemble 
$i$ and $n_r$ the number of ensembles. $n_r$ replicas are employed to sample 
this extended ensemble, each one placed at each ensemble. The definition of 
$Q_{ext}$ allows the introduction of swap trial moves between any two replicas, 
whenever the detail balance condition is satisfied. Since hard particles are 
being studied, it is convenient to expand isobaric-isothermal ensembles in 
pressure~\cite{Odriozola09}. This way, the partition function of the extended 
ensemble is given by~\cite{Okabe01,Odriozola09} 
\begin{equation}
Q_{\rm ext}=\prod_{i=1}^{n_r} Q_{N T P_i},
\end{equation} 
where $Q_{NTP_i}$ is the partition function of the isobaric-isothermal ensemble 
of the system at pressure $P_i$, temperature $T$, and with $N$ particles. 

The $NTP_i$ ensembles are sampled through a standard implementation, involving 
independent trial displacements, rotations of single ellipsoids, and volume 
changes. To increase the degrees of freedom of our relatively small systems ($N 
\sim 400$), we have implemented non-orthogonal parallelogram cells. Thus, sampling 
also includes trial changes of the angles and relative length sides of the cell 
lattice vectors. The probabilities for choosing any adjacent pairs of replicas 
are set equal and the following acceptance rule is employed~\cite{Odriozola09}
\begin{equation}
\label{accP} 
P_{\rm acc}\!=\! \min(1,\exp[\beta(P_i-P_j)(V_i-V_j)]), 
\end{equation} 
where $V_i-V_j$ is the volume difference between replicas $i$ and $j$ and 
$\beta=1/(k_BT)$ is the reciprocal temperature. Adjacent pressures must be close 
enough to provide reasonable swap acceptance rates. 

Simulations are started from a packed triangular arrangement of disks which are 
elongated in a certain in-plane direction by a factor $\kappa$. Conversely to 
the stretching of spheres, this procedure leads to the largest packed 
arrangement of ellipses~\cite{Donev04a}. It is 
faster to get a stationary state by decompressing packed cells than by 
compressing lose random configurations~\cite{Bautista13}. We first perform 
the necessary trial moves at the desired state points to ensure the 
development of a stationary state (on the order of $1 \times 10^{12}$ trial 
moves). During this process we adjust maximum displacements to get acceptance 
rates close to 0.3. We also relocate set pressures, initially set by following a 
geometric progression with the replica index, to obtain similar swap acceptance 
rates for all pairs of adjacent ensembles~\cite{Rathore05}. Once this is done, we 
then perform $4 \times 10^{12}$ additional sampling trials where maximum 
particle displacements, maximum rotational displacements, maximum volume 
changes, maximum changes of the lattice vectors, and pressures are fixed. Verlet 
neighbor lists~\cite{Donev05a} are used to improve performance. We set 
$N\sim400$ ellipsoids and $n_r=16$ or $32$, depending on the pressure range to 
be covered. $N\sim400$ seems enough in view of Xu et.~al.~analysis of system size effects~\cite{Xu13}. 
More details on the employed methods are given in previous works~\cite{Bautista13}.

\section{Results}

\begin{figure*}
\centering
\includegraphics[height=10cm]{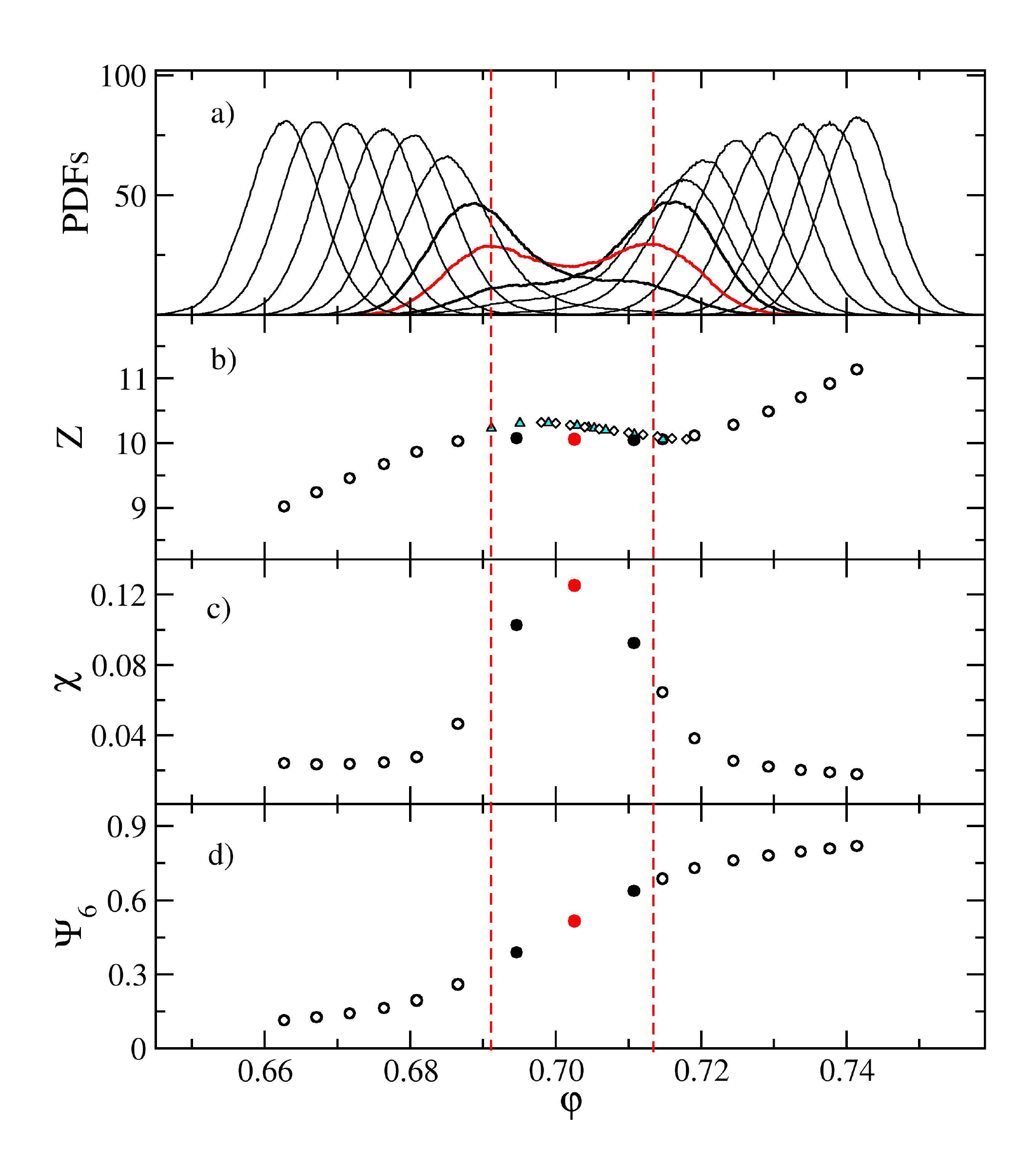}
\caption{\label{EOS-disks} Equation of state of hard disks at the fluid-solid 
transition. a) Probability density functions (PDFs) for all set pressures. The thick 
red line (bimodal) corresponds to the pressure at which the transition take 
place. Lines corresponding to the adjacent set pressures are also thicker than 
the others. b) Compressibility factor, $Z$, as a function of the area fraction, 
$\varphi$. c) Dimensionless isothermal compressibility, 
$\chi$. d) Overall bond order parameter, $\Psi_6$. Filled symbols of panels 
b)-d) correspond to the thicker lines of panel a). Dashed vertical lines are 
located at the peaks of the bimodal PDF. Small triangles and diamonds in panel 
b) are taken from the large scale simulations of Jaster ($N=128^2$) and Bernard and Krauth 
($N=1024^2$), respectively~\cite{Jaster99,Bernard11}. }
\end{figure*}

We start this section with the equation of state (EOS) for the $\kappa=1$ case, 
that is, for a system of disks on a plane. The EOS is shown in panel b) of 
figure \ref{EOS-disks}, i.~e.~$Z(\varphi)$ where $Z=\beta P/\rho$, 
$\varphi=a_e\rho$, $a_e=\pi\sigma_a\sigma_b/4=\pi\sigma_b^2\kappa/4$, 
$\rho=N/A$, and $A$ is the area of the simulation cell. In this plot we are also 
including as solid triangles the results of the large scale $NVT$ MC simulations with $N=128^2$ carried out by Jaster~\cite{Jaster99}, and as diamonds those of Bernard and Krauth~\cite{Bernard11} with $N=1024^2$. These results are set in the liquid-solid 
transition region where size effects are expected to be present. As can be seen, 
differences with our data are not very large, though. Nonetheless, even very large systems show small differences with increasing the system size~\cite{Bernard11}. In addition to the EOS, we are 
including the probability density functions (PDFs) from where the averages are 
taken (panel a), the dimensionless isothermal compressibility 
$\chi=N(\langle\rho^2\rangle-\langle\rho\rangle^2)/\langle\rho\rangle^2$ (panel 
c), and the global order parameter $\Psi_6=1/N|\sum_i^N\varphi_{6,i}|$ (panel d) 
with $\varphi_{6,i}=1/N_i^b\sum_j^{N_i^b} \exp(6\theta_{ij}\sqrt{-1})$ where 
$N_i^b$ is the number of bonding particles to $i$ and $\theta_{ij}$ is the angle 
between the $ij$-bond and an arbitrary fixed reference axis. All these data 
strongly suggest a first order transition. Nevertheless, the $Z(\varphi)$-plateau, the 
$\chi(\varphi)$ peak, and the development of an overall bond order are well 
known facts which do not constitute enough evidence to establish the nature of the liquid-solid transition. 
Indeed, there is not a general consensus on the nature of the fluid-solid hard disks 
transition. Among the different scenarios, the KTHNY theory~\cite{Kosterlitz-Thouless,Halperin-Nelson} predicts a two-step 
transition where the fluid turns into a hexatic phase before the solid when 
increasing pressure. According to the KTHNY theory both transitions, i.~e.~fluid-hexatic and hexatic-solid, are continuous. Recent large scale ($N=1024^2$) computer simulations, however, support the existence of a first order 
fluid-hexatic transition followed by a hexatic-solid continuous one~\cite{Bernard11} (a bubble formation, which is a hallmark of a first-order transition, is observed). This work reports a coexistence interval 
of $0.700<\varphi<0.716$ for the fluid-hexatic transition and a second transition at 
$\varphi \gtrsim 0.720$. Hence, the hexatic phase would only take place at the interval $0.716<\varphi<0.720$. 

Back to our results, we did obtain a PDF bimodal at the coexistence region. The 
curve is highlighted in panel a) of figure~\ref{EOS-disks}. This bimodal also 
supports the existence of a first order transition. From their peaks we obtain the coexistence region
$0.691\lesssim \varphi \lesssim 0.713$ (we follow the 
histogram reweighting technique for determining the coexistence 
boundaries~\cite{Ferrenberg88,Ferrenberg89}). The vertical dashed lines of figure~\ref{EOS-disks} point 
out the coexistence interval. Taking into account Bernard and Krauth 
conclusions, this coexistence should be fluid-hexatic. The hexatic-solid would 
be relatively close to $\varphi_{s} \approx 0.717$, but we are not capturing 
this subtle continuous transition (Bernard and Krauth capture it from the shift 
of positional order decay from exponential to power-law, on a length scale of 
$\sim 100 \sigma_b$). It should also be noted that 
our coexistence region is wider and shifted to the left as compared to the 
fluid-hexatic coexistence given in reference~\cite{Bernard11}. This not so large 
mismatch is a consequence of finite size effects.       

\begin{figure*}
\centering
\includegraphics[height=10cm]{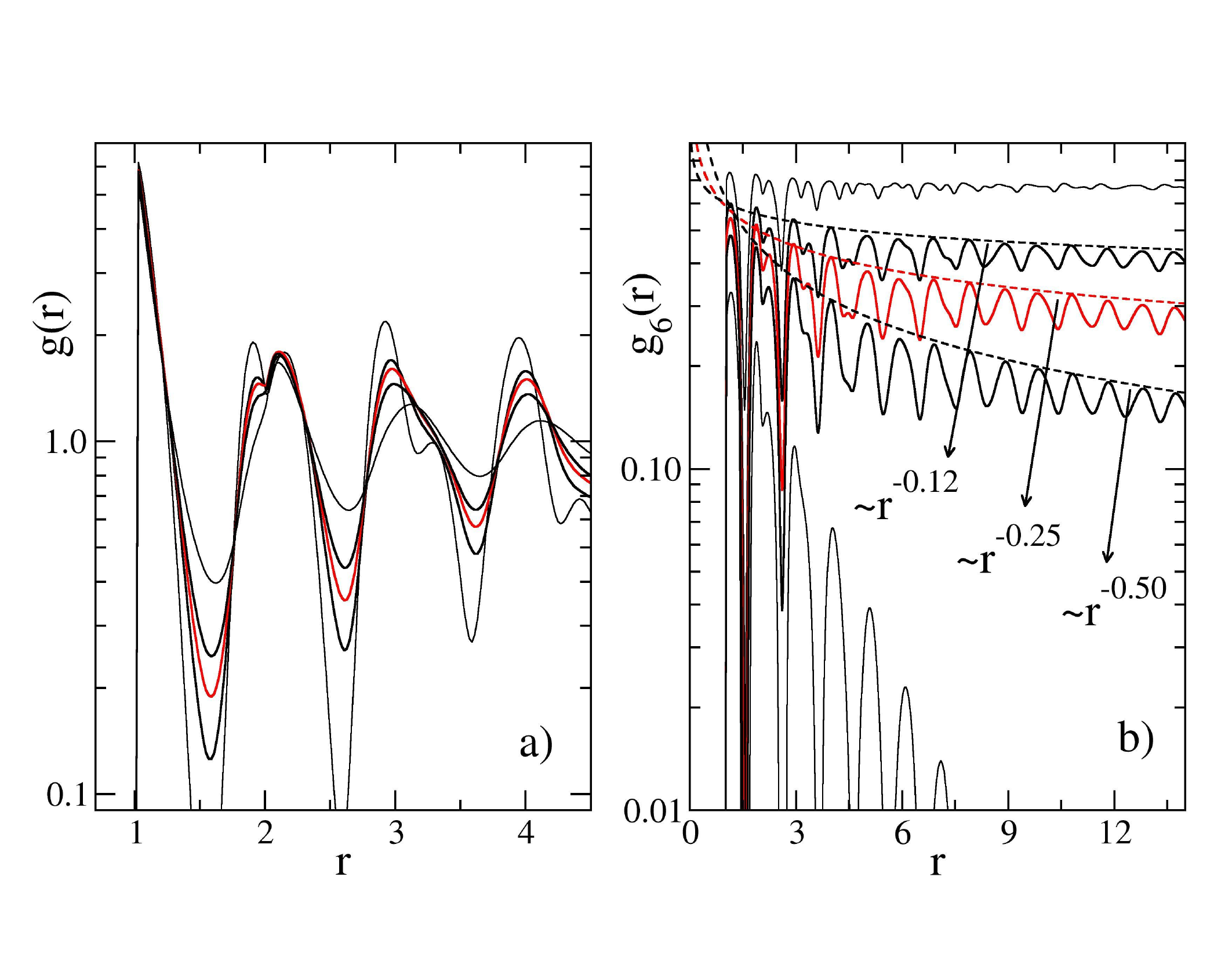}
\caption{\label{gs(r)-k11} a) Radial distribution functions, $g(r)$, for a 
system of disks at the pressure of the fluid-hexatic transition (red thick 
line), adjacent set pressures (black thick lines), and highest and lowest set 
pressures (black thin lines). Thick lines correspond to the filled symbols of 
figure~\ref{EOS-disks}. b) The corresponding bond orientation radial correlation 
functions, $g_6(r)$, for the same pressures. The dashed lines point out the 
$g_6(r)$ decay with distance.}
\end{figure*}

Letting aside the peaks of the PDF bimodal, the bond orientation radial 
correlation function is frequently employed to quantitatively identify the location 
of the transition from the isotropic liquid to the hexatic phase~\cite{Xu13}. It is given by $g_6(r)=\langle\sum_{i\neq 
j}\delta(r-r_{ij})\varphi_{6,i}\varphi_{6,j}^*/\sum_{i\neq 
j}\delta(r-r_{ij})\rangle$, where $\delta$ is the Kronecker delta function. The hexatic phase sets in when the $g_6(r)$ function decays slower than $g_6(r) \sim r^{-\eta_6}$ with $\eta_6=1/4$ as pointed out elsewhere~\cite{Kosterlitz-Thouless,Halperin-Nelson}. We are showing the obtained $g_6(r)$ curves in panel b) of figure~\ref{gs(r)-k11} as obtained from the bimodal PDF shown in figure~\ref{EOS-disks} and both adjacent set pressures. We are also including the $g_6(r)$ curves for the highest and lowest set pressures as thin lines. As 
labeled in panel b), the thick dashed lines correspond to $\sim r^{-\eta}$ with 
$\eta=0.12$, 0.25, and 0.50. We estimate these values to have relatively large 
error bars (though below 20$\%$) due to the small system size we are employing. 
Note that the $\sim r^{-\eta_6}$ decay matches the overall $g_6(r)$ trend 
obtained from the bimodal. Hence, both criteria, the 
double-peak interval from the PDF bimodal and the $g_6(r) \sim r^{-\eta_6}$ trend 
coincide for the location of the liquid-hexatic boundary. We also observe that 
$\Psi_6(\varphi)$ shows an inflection at this point (panel d) of figure~\ref{EOS-disks}). We are following the double-peak criteria for the phase diagram construction. In panel a) of the same figure we are including the 
corresponding radial distribution functions, $g(r)$. As expected, the structure 
builds up with increasing pressure, when the centers of mass of the particles arrange in a triangular lattice. Nonetheless, the positional correlations exponentially decay on a length scale of $\sim 100 \sigma_b$ in the hexatic phase~\cite{Bernard11}.

\begin{figure*}
\centering
\includegraphics[height=10cm]{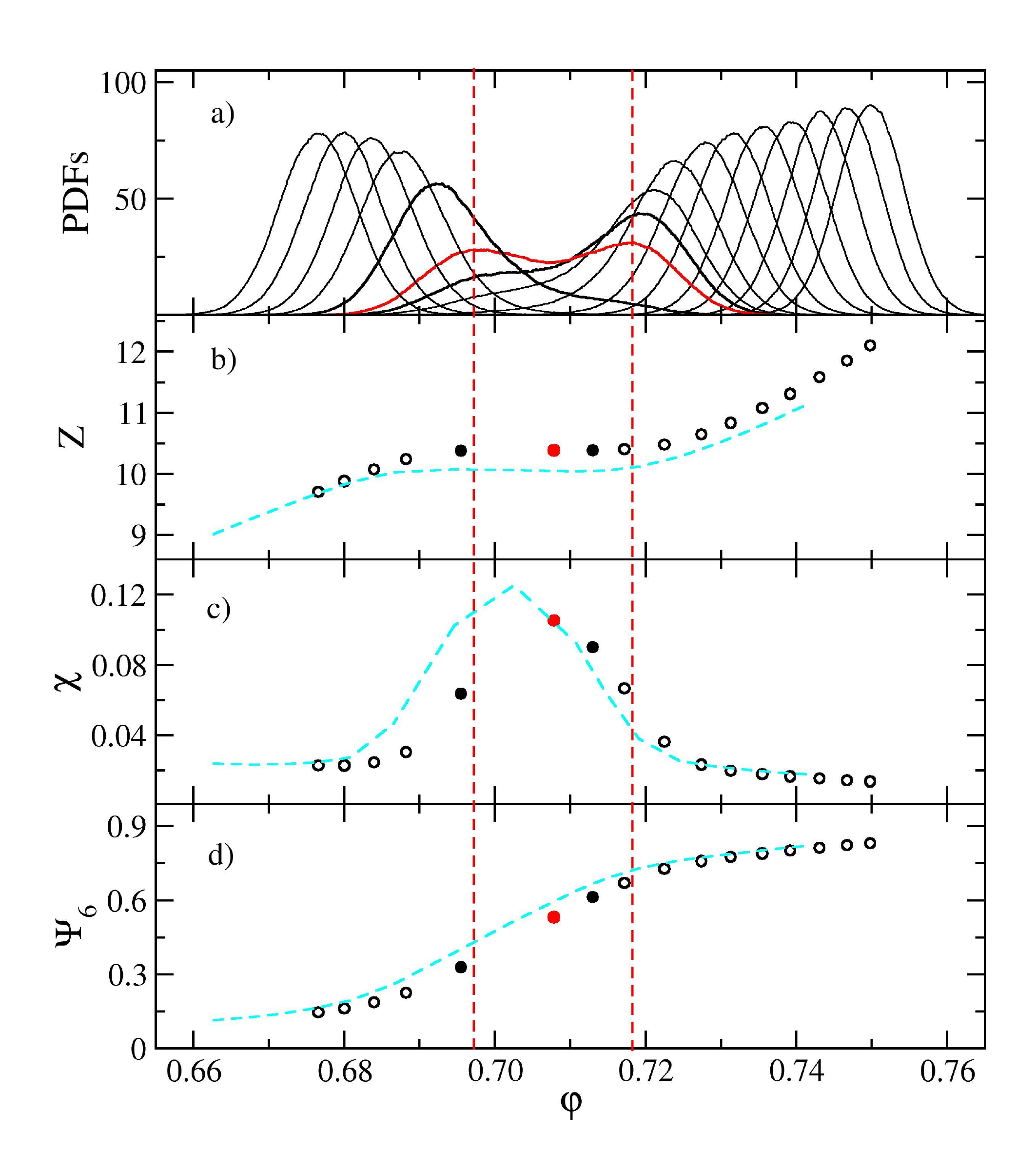}
\caption{\label{EOS-k11} Equation of state for $\kappa=1.1$ at the 
isotropic-plastic transition. a) Probability density functions for each set 
pressure. The thick red line (bimodal) corresponds to the pressure at which the 
transition take place. Lines corresponding to the adjacent set pressures are 
also thicker than the others. b) Compressibility factor, $Z$, as a function of 
the area fraction, $\varphi$. c) Dimensionless isothermal compressibility, $\chi$. d) Overall bond order parameter, $\Psi_6$. 
Filled symbols of panels b)-d) correspond to the thicker lines of panel a). 
Dashed vertical lines are located at the peaks of the PDF bimodal. Cyan (light) 
dotted lines in panels b)-d) show the data for disks.}
\end{figure*}

The equation of state for $\kappa=1.1$ is shown in panel b) of figure~\ref{EOS-k11} for the isotropic-plastic transition. As for the disks case, we are showing the probability density functions (PDFs) (panel a), the 
dimensionless isothermal compressibility $\chi(\varphi)$ (panel c), and the global order parameter $\Psi_6(\varphi)$ 
(panel d). In panels b)-d) we include as cyan dashed lines the results 
obtained for $\kappa=1.0$ to make the comparison easy. From all panels it 
becomes clear that the transition slightly shifted to higher densities and that 
the coexistence region narrows. In addition, the transition pressure also 
increases. All this is a consequence of the smaller entropy gain associated 
to the transition, since ellipses at the plastic phase do not pack as well as 
disks do~\cite{Vieillard-Baron72}. Furthermore, the shifting to higher densities and narrowing of the coexistence 
region continues for increasing $\kappa$. This trend remains up to $\kappa 
\lesssim 1.6$, where the plastic region vanishes. We take this end to define the 
upper point of the low anisotropy region. This behavior is similar to those 
observed for spheroids, although the limit for the plastic region in this case 
is around 1.33 (for both sides, prolates and oblates)~\cite{Bautista13}. 
Finally, we add here that for all studied $\kappa$ our EOSs perfectly match 
those recently reported by Xu et.~al.~\cite{Xu13} (not shown). Hence, their 
conclusions on the validity of several theoretical EOS~\cite{Varga98,Boublik11} 
remain unchanged when considering our results. However, as shown further in the 
text, we are pressurizing the system as much as necessary to access the solid 
phase.

\begin{figure*}
\centering
\includegraphics[height=10cm]{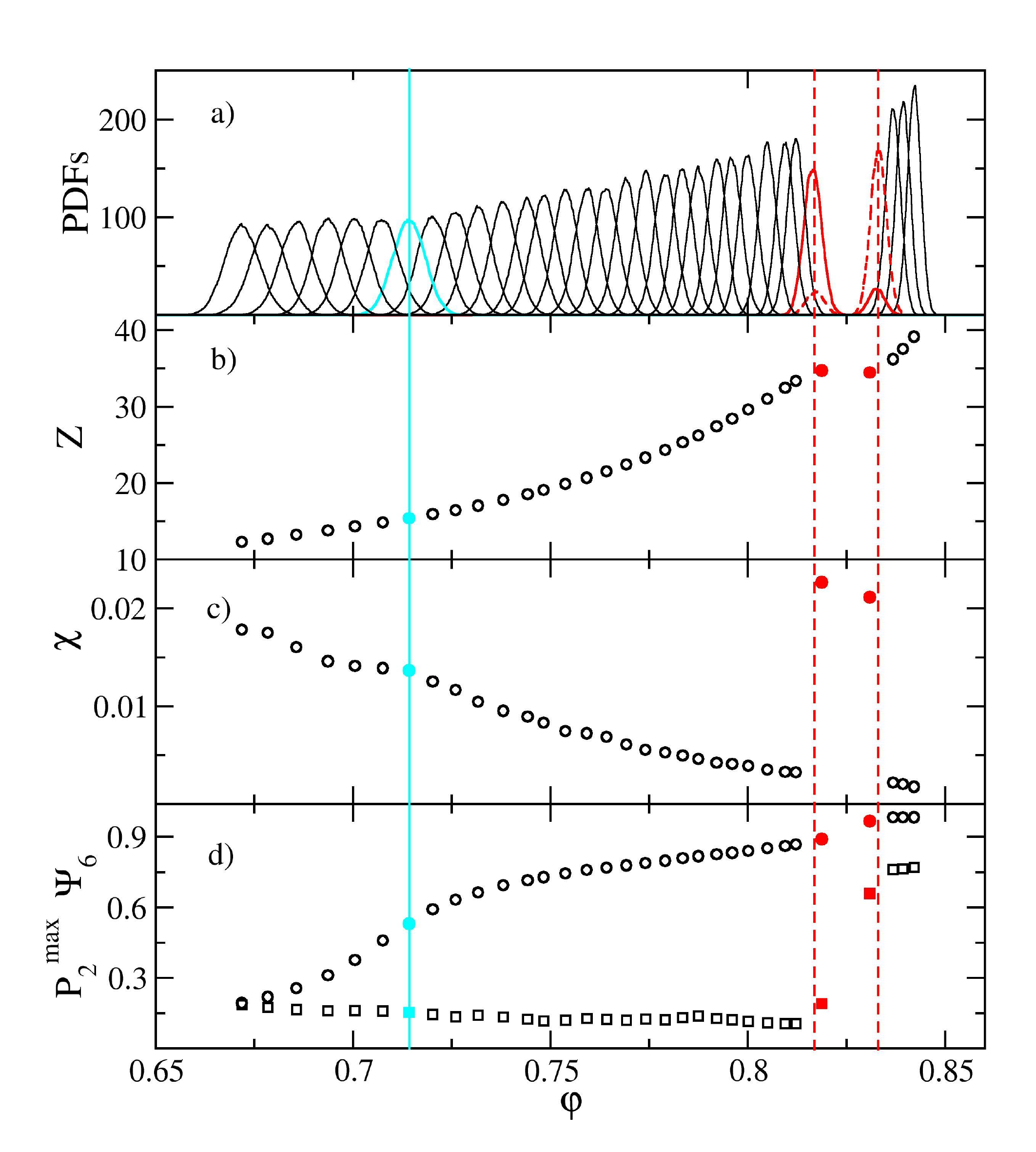}
\caption{\label{EOS-k35} Equation of state for $\kappa=3.5$. a) Probability 
density functions for each set pressure. b) Compressibility factor, $Z$, as a 
function of the area fraction, $\varphi$. c) Dimensionless isothermal compressibility, $\chi$. d) Overall orientational order parameter, $P_2^{max}$ (circles), and the global bond-order parameter $\Psi_6$ (squares). The space between the vertical dashed lines corresponds to the nematic-solid coexistence region. The vertical solid line points out the isotropic-nematic continuous transition.}
\end{figure*}


Figure~\ref{EOS-k35} is an example of the EOS we have obtained for the large 
anisotropy region, $\kappa \gtrsim 2.4$. In particular, this figure is built for 
$\kappa=3.5$. Again, the PDFs, $Z(\varphi)$, $\chi(\varphi)$, 
$P_2^{max}(\varphi)$, and $\Psi_6(\varphi)$ are shown in the panels. $P_2^{max}$ 
is the overall orientational order parameter (or nematic order parameter) which 
is given by the maximum eigenvalue of the tensor order 
parameter~\cite{Cuesta90}. In 2D $P_2^{max}=[\langle 1/N \sum_i^N 
\cos(2\theta_i) \rangle^2 + \langle 1/N \sum_i^N \sin(2\theta_i) 
\rangle^2]^{1/2}$, where $\theta_{i}$ is the angle between 
$\hat{\mathbf{u}}_{i}$ and an arbitrary fixed direction. On the other hand, the 
angle between the nematic director and the same arbitrary direction is 
$\theta_{dir}=\tan^{-1}[(P_2^{max}-\langle 1/N \sum_i^N \cos(2\theta_i) 
\rangle)/ \langle 1/N \sum_i^N \sin(2\theta_i) \rangle]$. Alternatively, 
$P_2^{max}$ can be numerically obtained as described elsewhere~\cite{Xu13}. 
Although $P_2^{max}$ decreases with the system size for the quasi-long-range 
nematic phase, its dependence is not strong~\cite{Xu13}. Here two transitions 
are detected. A continuous one, at low compressions, which corresponds to an 
isotropic-nematic transition; and another that is discontinuous, at higher 
pressures, corresponding to a nematic-solid transition. No further transitions 
where observed at larger densities. The first transition is characterized by an 
increase of $P_2^{max}(\varphi)$, an invariant $\Psi_6(\varphi)$, a tiny plateau 
of $Z(\varphi)$, and a small bump of $\chi(\varphi)$. We are locating the 
isotropic-nematic transition at this bump. The PDF corresponding to this 
transition is clearly monomodal and supports a disclination unbinding scenario. 
We indeed obtain the same result for all anisotropies above $2.5$. This result 
contradicts the Cuesta and Frenkel claim~\cite{Cuesta90} that the 
isotropic-nematic transition is first order for $\kappa=4$. Hence, according to 
our data, there is no tricritical point on the isotropic-nematic transition 
line. At the second transition we find an increase of $P_2^{max}(\varphi)$, a 
steep jump of $\Psi_6(\varphi)$, a large plateau of $Z(\varphi)$, and an 
important bump of $\chi(\varphi)$. All these signatures appear together with the 
PDFs bimodals at the coexistence region. Note that for $\kappa>2.0$ we are 
applying an stretching procedure with a factor $\kappa$ to all bonds along the 
director direction previous to the $\Psi_6$ computation. This is done in order 
to obtain $\Psi_6=1$ for a perfect crystal and to take advantage of the $\Psi_6$ 
definition.    

\begin{figure*}
\centering
\includegraphics[height=10cm]{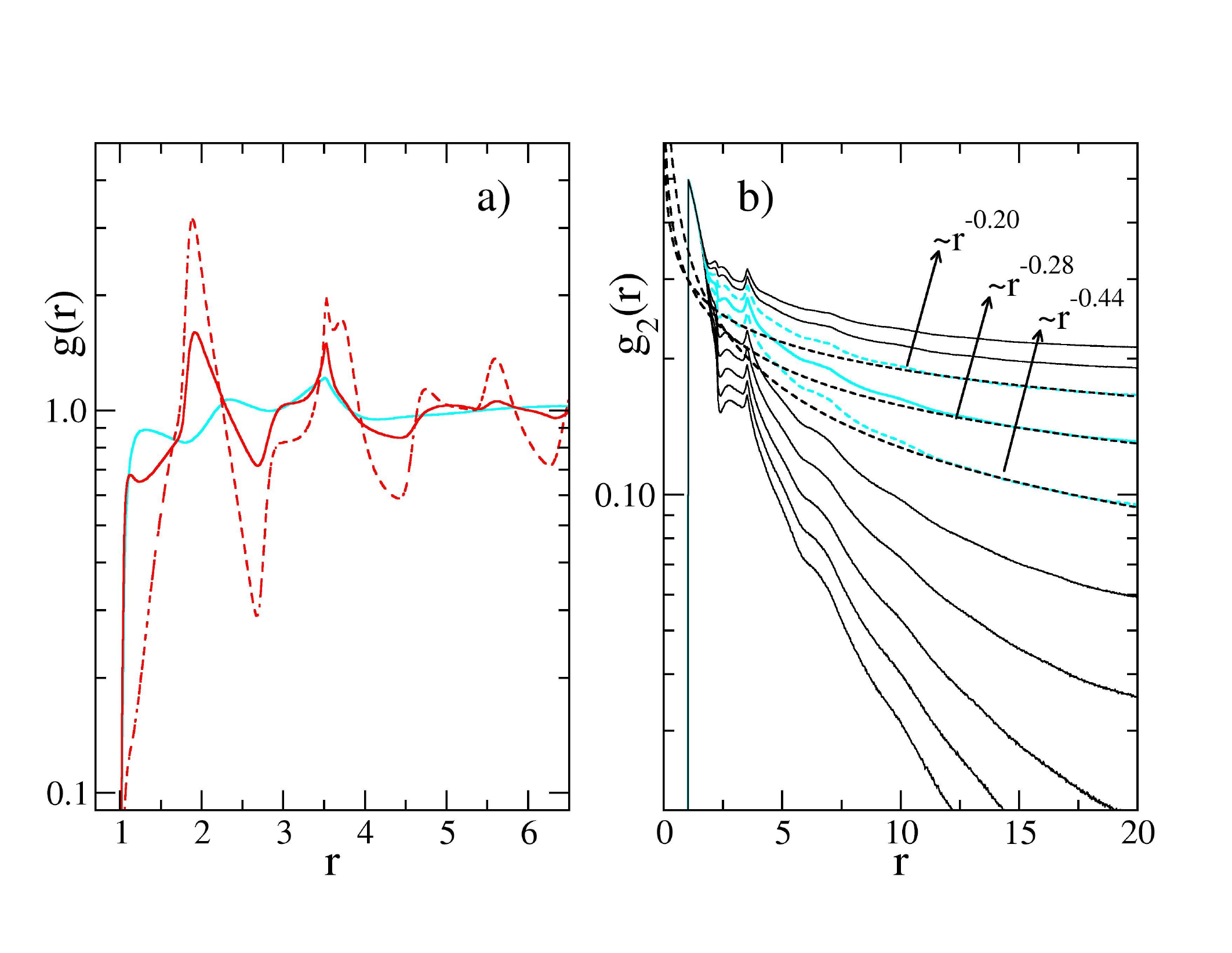}
\caption{\label{gs(r)-k35} a) Radial distribution functions, $g(r)$, for 
$\kappa=3.5$ at the area fraction of the isotropic-nematic (cyan 
solid line) and nematic-solid (red solid and dashed lines) transitions. b) 
Radial orientational functions, $g_2(r)$, above and below the isotropic-nematic transition. The thick solid (cyan) line corresponds to the 
isotropic-nematic pressure whereas the thick dashed (cyan) lines to the adjacent pressures. Dashed black lines correspond to the quasi-long-range $g_2(r) \sim r^{-\nu}$ fit. }
\end{figure*}

An alternative way for determining an upper bound to the exact value of the 
isotropic-nematic pressure is by means of analyzing the decay of the angular 
correlation function~\cite{Straley71,Frenkel-Eppenga,Bates-Frenkel,Xu13}, 
$g_2(r)$. The 2D nematic phase is characterized by a power-law decay of 
$g_2(r)\sim r^{-\eta}$ with $\eta < \eta_2=1/4$. This algebraic decay is a 
common feature for needles, rods, and ellipses confined to a plane. Thus, the 
subensemble average at the smallest pressure which leads to $\eta < 1/4$ can be 
considered to produce the 2D-nematic phase the closest to the isotropic phase. 
The fittings of $g_2(r)~r^{-\eta}$ for the curves obtained in the vicinity of 
the isotropic-nematic pressure are shown in panel b) of figure~\ref{gs(r)-k35}. 
As labeled, we get $\eta=0.28$ for the transition determined according to the 
$\chi(\varphi)$ bump criterion. So, the $g_2(r)$ fit analysis give 
rises to a slightly larger area fraction value for this transition than that 
obtained from the $\chi(\varphi)$ bump. This result is observed for all set 
anisotropies in the large anisotropy region. For completeness, panel a) of the 
same figure shows the radial distribution function obtained for the 
isotropic-nematic transition (cyan solid line) and those obtained before and 
after the nematic-solid transition (red dashed and solid lines).   

\begin{figure*}
\centering
\includegraphics[height=10cm]{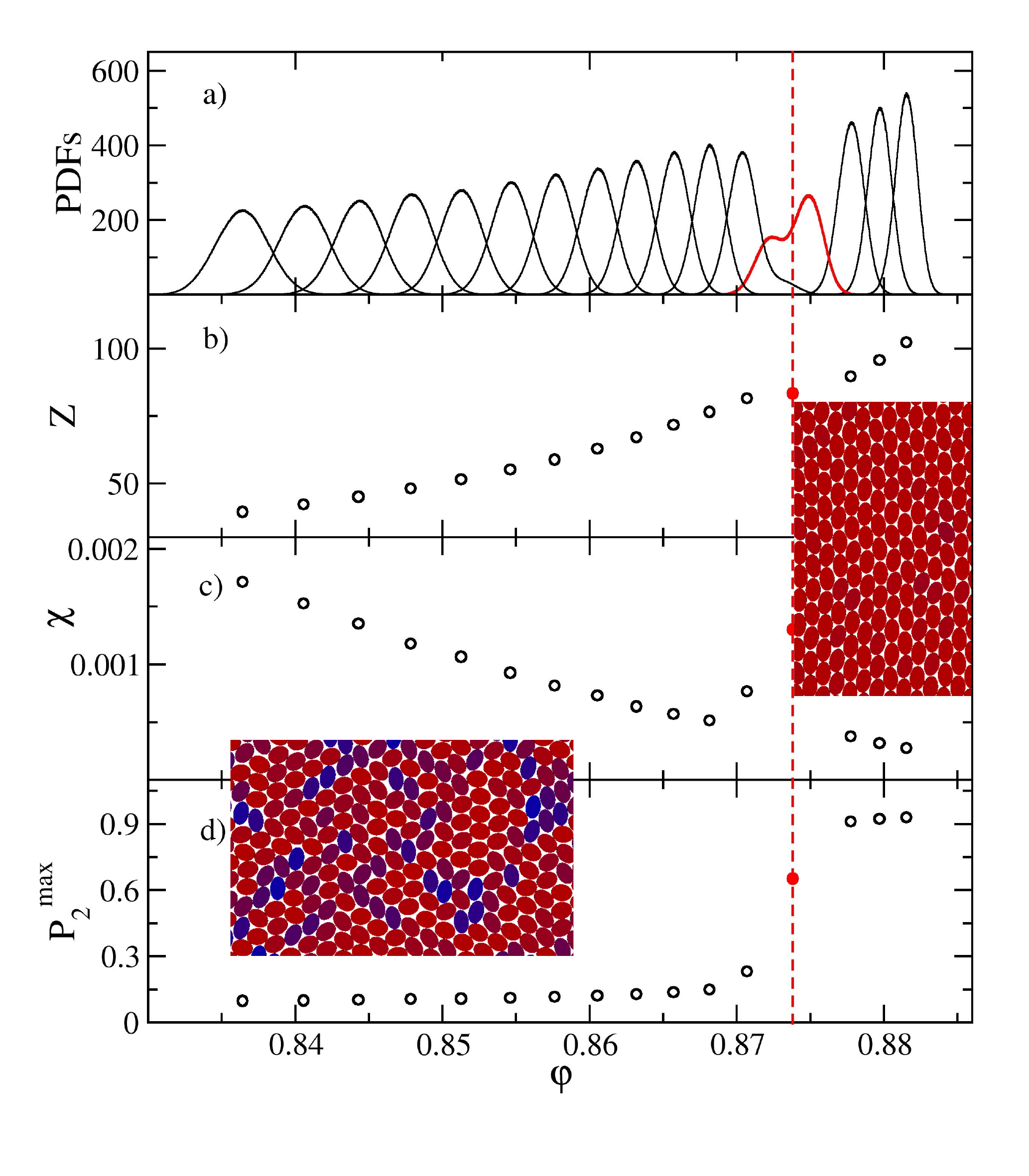}
\caption{\label{EOS-k14-alta} Equation of state for $\kappa=1.4$ at high 
densities (plastic-solid transition). a) Probability density functions for all  
set pressures. b) Compressibility factor, $Z$, as a function of the area 
fraction, $\varphi$. c) The dimensionless isothermal 
compressibility, $\chi$. d) Overall orientational order parameter, $P_2^{max}$. 
The dashed (red) line and filled symbols point out the results for the pressure 
at which the transition takes place. }
\end{figure*}

Up to this point, we have focused on the isotropic-nematic and nematic-solid 
transitions for large anisotropies, and on the isotropic-plastic transition for 
low anisotropies occurring at relatively low pressures. Hence, the high-pressure 
plastic-solid transition is still missing. In order to capture this transition, 
we decompress perfect crystal cells with $\kappa\leq1.5$ at high pressures. In 
particular, we are showing in figure~\ref{EOS-k14-alta} the results obtained for 
$\kappa=1.4$. In this plot we are including the PDFs, $Z(\varphi)$, 
$\chi(\varphi)$, and $P_2^{max}(\varphi)$. We are also including a couple of 
snapshots of part of the system cells showing the plastic (left) and solid 
(right) phases. As shown in panel b) of figure~\ref{EOS-k14-alta} a bimodal PDF 
curve builds up at $\varphi \approx 0.874$ (red and thick solid line). The 
histogram reweighting procedure~\cite{Ferrenberg88,Ferrenberg89} leads to a 
valley at $\varphi \approx 0.873$ (we are taking this point for the phase 
diagram). This curve corresponds to the solid symbols appearing at the other 
panels. Thus, for this PDF we observe a $Z(\varphi)$ plateau, a $\chi(\varphi)$ 
bump, and a steep increase of $P_2^{max}(\varphi)$. These features suggest a 
discontinuous transition and a small coexistence region. Nonetheless, the nature 
of the transition turns unclear for decreasing $\kappa$, as the bimodals seem to 
disappear, producing small kinks for $\chi(\varphi)$ (not shown).          

\begin{figure*}
\centering
\includegraphics[height=10cm]{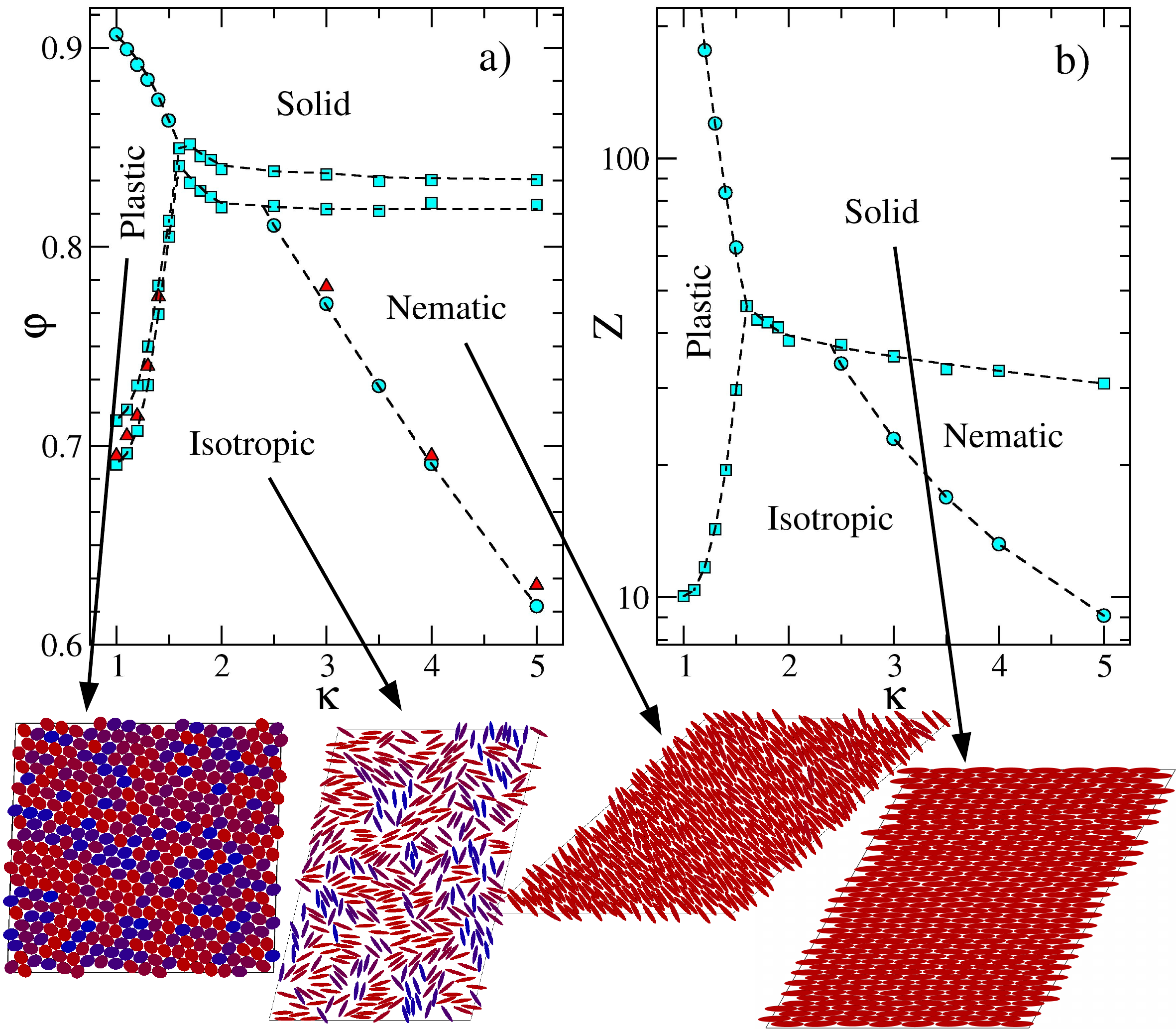}
\caption{\label{phase-diagram} a) Phase diagram of hard ellipses. The disks 
case is given for $\kappa=1$. There are several transition types. These are: 
isotropic-plastic, isotropic-solid, isotropic-nematic, nematic-solid, and 
plastic-solid. Square pairs are employed to point out the limits of first order 
transitions and single circles are used for continuous transitions (the only 
exceptions to this notation are the plastic-solid transition for $\kappa=1.4$ 
and $1.5$, which are found to be discontinuous). b) The corresponding 
compressibility factor, $Z$, for the transitions. In both panels the dashed 
lines are guides to the eye. The snapshots correspond to the different regions 
of the phase diagram. These are, from left to right, plastic for $\kappa=1.4$ 
and isotropic, nematic, and solid for $\kappa=5.0$. Note that red is employed 
for particles aligned with the nematic director and blue for those perpendicular 
to it. A linear combination of both colors is employed for intermediate cases.  
}
\end{figure*}

The phase diagram of hard ellipses is build by gathering the information for 
all studied $\kappa$. This is shown in the left panel of 
figure~\ref{phase-diagram}. At the right panel of the same figure we are 
including the compressibility factor, $Z$, at which the transitions take place. 
Furthermore, snapshots corresponding to the different phases are embedded in the 
figure. There we can see the plastic, isotropic, nematic, and solid phases. We 
are using a couple of square symbols to point out a coexistence region 
(in-between the couple) and single circles to point out a continuous transition. 
Note that we are marking the plastic-solid transition with circles, though a 
tiny coexistence is found for $\kappa=1.4$ and 1.5. The dashed lines are guides 
to the eye. We are also including, as red triangles, the recently published data 
by Xu et.~al.~\cite{Xu13}. As can be seen, our results well agree with their 
predictions for both, the isotropic-plastic and the isotropic-nematic 
transitions. A comparison with data from Cuesta and Frenkel~\cite{Cuesta90} and 
Vieillard-Baron~\cite{Vieillard-Baron72} is provided in reference~\cite{Xu13}. 

The phase diagram of hard ellipses can be split into three regions. For low 
anisotropies, $1 < \kappa \lesssim 1.6$, there are two transitions. A 
low-pressure isotropic-plastic and a high-pressure plastic-solid transition. The 
first one, for $\kappa=1$ and according to Bernard and Krauth 
findings~\cite{Bernard11}, is an isotropic-hexatic first order transition 
followed by a subtle hexatic-solid transition. Consequently, for $1<\kappa 
\lesssim 1.6$ we obtain an isotropic-hexatic first order transition, which is 
the one we are capturing, followed by a mild hexatic-plastic continuous 
transition, which we are not detecting. The same conclusion is supported by the 
$g_6(r)$ analysis given elsewhere~\cite{Xu13}. Since the hexatic region is tiny, 
we are not including it in the phase diagram. The high pressure transition is 
also first order, at least for $\kappa=1.4$ and 1.5. The second region 
corresponds to intermediate anisotropies, i.~e. for $1.6 \lesssim \kappa 
\lesssim 2.4$. Here it is observed only a single isotropic-solid transition, 
where both, bond-orientational and orientational order develop. Finally, the 
third region corresponds to $\kappa \gtrsim 2.4$, where an isotropic-nematic 
transition occurs at low pressure and a nematic-solid transition appears at high 
pressure. This last transition is observed above an area fraction of $0.8$ for 
all $\kappa$. This quantitative result differs from those reported in 
references~\cite{Vieillard-Baron72,Cuesta90} but agrees with Xu et.~al.~recent 
results~\cite{Xu13} (they found no sign of a transition involving a solid below 
$\varphi=0.8$). We point out the weak dependence of the nematic-solid transition 
on $\kappa$. That is, it occurs at an almost constant area fraction and pressure 
(dependence on the pressure is larger, though). This suggests that the 
nematic-solid entropy gain associated to the transition practically holds with 
increasing $\kappa$, which in turn implies a similar gain on the system 
accessible area. In other words, the system behaves like being stretched in the 
nematic director direction while preserving occupied, accessible, and excluded 
areas.  

\section{Conclusions}
We have reported the phase diagram of hard ellipses for anisotropies in the 
range $1\leq \kappa \leq 5$. This is done by means of replica exchange Monte 
Carlo simulations. For $1\leq \kappa \lesssim 1.6$ we have found an isotropic 
phase at low pressures, a plastic one at intermediate pressures, and a solid one 
at high-pressures. In this case, the isotropic-plastic transition would probably 
be a two-step transition, with a small hexatic phase region in-between the 
isotropic and plastic regions. Our data support the existence of a first order 
transition in agreement with large-scale simulations of disks~\cite{Bernard11}. 
In addition, the high pressure transition (plastic-solid) close to the upper 
bound of $\kappa$ is also discontinuous. For weak anisotropies we have obtained 
a plastic-solid continuous transition. This would imply a tricritical point 
somewhere in the range $1.2 \lesssim \kappa \lesssim 1.4$. This picture, 
however, may probably change when considering larger system sizes in favor of 
the discontinuous scenario. For intermediate anisotropies, $1.6 \lesssim \kappa 
\lesssim 2.4$, the system shows only a single first order transition, 
isotropic-solid. Thus, nematic is absent here, in agreement with Cuesta and 
Frenkel early results~\cite{Cuesta90}. Finally, for $\kappa \gtrsim 2.4$, a 
continuous isotropic-nematic and a discontinuous nematic-solid transition are 
found. Our reported boundaries for the anisotropy regions are slightly different 
from those recently reported by Xu et.~al.~\cite{Xu13}. These differences mostly 
appear due to the fact that their study does not include results for area 
fractions above $0.8$.

Finally, we think it is worth mentioning some similarities of hard 
anisotropic objects of variable aspect ratio between the 2D and 3D scenarios. 
One should note that the overall appearance obtained for the 2D phase diagram 
(Fig.~\ref{phase-diagram}) markedly resembles that of 3D systems of prolate and 
oblate ellipsoids, spherocylinders, and cut-spheres with variable aspect ratio 
(see references~\cite{Bolhuis97,Wensink09,Marechal11,Bautista13}). In 
particular, the isotropic-nematic transition line goes up in occupied area 
(volume) fraction upon decreasing the particle aspect ratio to eventually meet 
up with a strongly first-order and almost anisometric-independent fluid-solid 
transition. This point defines a critical aspect ratio below 
which the nematic phase ceases to be thermodynamically stable. This is a common 
feature for all referenced systems and most probably for other  
convex particle shapes in two and three dimensions. 

\begin{acknowledgments}
The authors thank Prof.~Eliezer Braun for his fruitful discussions and support,
Wen-Sheng Xu, Yan-Wei Li, Zhao-Yan Sun, and Li-Jia An for sharing their EOS 
data, Szabolcs Varga for useful suggestions, and the reviewers for their helpful 
comments. Indeed, the closing paragraph is taken from one of them. G.B-C thanks 
CONACyT for a Phd.~scholarship. G.O thanks CONACyT Project No.~169125 for 
financial support.
\end{acknowledgments}


\begin{thebibliography}{47}%
\makeatletter
\providecommand \@ifxundefined [1]{%
 \@ifx{#1\undefined}
}%
\providecommand \@ifnum [1]{%
 \ifnum #1\expandafter \@firstoftwo
 \else \expandafter \@secondoftwo
 \fi
}%
\providecommand \@ifx [1]{%
 \ifx #1\expandafter \@firstoftwo
 \else \expandafter \@secondoftwo
 \fi
}%
\providecommand \natexlab [1]{#1}%
\providecommand \enquote  [1]{``#1''}%
\providecommand \bibnamefont  [1]{#1}%
\providecommand \bibfnamefont [1]{#1}%
\providecommand \citenamefont [1]{#1}%
\providecommand \href@noop [0]{\@secondoftwo}%
\providecommand \href [0]{\begingroup \@sanitize@url \@href}%
\providecommand \@href[1]{\@@startlink{#1}\@@href}%
\providecommand \@@href[1]{\endgroup#1\@@endlink}%
\providecommand \@sanitize@url [0]{\catcode `\\12\catcode `\$12\catcode
  `\&12\catcode `\#12\catcode `\^12\catcode `\_12\catcode `\%12\relax}%
\providecommand \@@startlink[1]{}%
\providecommand \@@endlink[0]{}%
\providecommand \url  [0]{\begingroup\@sanitize@url \@url }%
\providecommand \@url [1]{\endgroup\@href {#1}{\urlprefix }}%
\providecommand \urlprefix  [0]{URL }%
\providecommand \Eprint [0]{\href }%
\providecommand \doibase [0]{http://dx.doi.org/}%
\providecommand \selectlanguage [0]{\@gobble}%
\providecommand \bibinfo  [0]{\@secondoftwo}%
\providecommand \bibfield  [0]{\@secondoftwo}%
\providecommand \translation [1]{[#1]}%
\providecommand \BibitemOpen [0]{}%
\providecommand \bibitemStop [0]{}%
\providecommand \bibitemNoStop [0]{.\EOS\space}%
\providecommand \EOS [0]{\spacefactor3000\relax}%
\providecommand \BibitemShut  [1]{\csname bibitem#1\endcsname}%
\let\auto@bib@innerbib\@empty
\bibitem [{\citenamefont {Loewen}(2009)}]{Loewen09}%
  \BibitemOpen
  \bibfield  {author} {\bibinfo {author} {\bibfnamefont {H.}~\bibnamefont
  {Loewen}},\ }\href@noop {} {\bibfield  {journal} {\bibinfo  {journal} {J.
  Phys.: Condens. Matter}\ }\textbf {\bibinfo {volume} {21}},\ \bibinfo {pages}
  {474203} (\bibinfo {year} {2009})}\BibitemShut {NoStop}%
\bibitem [{\citenamefont {Tkalec}\ and\ \citenamefont
  {Muševič}(2013)}]{Tkalec13}%
  \BibitemOpen
  \bibfield  {author} {\bibinfo {author} {\bibfnamefont {U.}~\bibnamefont
  {Tkalec}}\ and\ \bibinfo {author} {\bibfnamefont {I.}~\bibnamefont
  {Muševič}},\ }\href@noop {} {\bibfield  {journal} {\bibinfo  {journal}
  {Soft Matter}\ }\textbf {\bibinfo {volume} {9}},\ \bibinfo {pages} {8140}
  (\bibinfo {year} {2013})}\BibitemShut {NoStop}%
\bibitem [{\citenamefont {Arciniegas}\ \emph {et~al.}(2014)\citenamefont
  {Arciniegas}, \citenamefont {Kim}, \citenamefont {De~Graaf}, \citenamefont
  {Brescia}, \citenamefont {Marras}, \citenamefont {Miszta}, \citenamefont
  {Dijkstra}, \citenamefont {van Roij},\ and\ \citenamefont
  {Manna}}]{Arciniegas14}%
  \BibitemOpen
  \bibfield  {author} {\bibinfo {author} {\bibfnamefont {M.~P.}\ \bibnamefont
  {Arciniegas}}, \bibinfo {author} {\bibfnamefont {M.~R.}\ \bibnamefont {Kim}},
  \bibinfo {author} {\bibfnamefont {J.}~\bibnamefont {De~Graaf}}, \bibinfo
  {author} {\bibfnamefont {R.}~\bibnamefont {Brescia}}, \bibinfo {author}
  {\bibfnamefont {S.}~\bibnamefont {Marras}}, \bibinfo {author} {\bibfnamefont
  {K.}~\bibnamefont {Miszta}}, \bibinfo {author} {\bibfnamefont
  {M.}~\bibnamefont {Dijkstra}}, \bibinfo {author} {\bibfnamefont
  {R.}~\bibnamefont {van Roij}}, \ and\ \bibinfo {author} {\bibfnamefont
  {L.}~\bibnamefont {Manna}},\ }\href@noop {} {\bibfield  {journal} {\bibinfo
  {journal} {Nano Letters}\ }\textbf {\bibinfo {volume} {14}},\ \bibinfo
  {pages} {1056} (\bibinfo {year} {2014})}\BibitemShut {NoStop}%
\bibitem [{\citenamefont {Quan}\ and\ \citenamefont {Fang}(2010)}]{Zewei10}%
  \BibitemOpen
  \bibfield  {author} {\bibinfo {author} {\bibfnamefont {Z.}~\bibnamefont
  {Quan}}\ and\ \bibinfo {author} {\bibfnamefont {J.}~\bibnamefont {Fang}},\
  }\href@noop {} {\bibfield  {journal} {\bibinfo  {journal} {Nano Today}\
  }\textbf {\bibinfo {volume} {5}},\ \bibinfo {pages} {390} (\bibinfo {year}
  {2010})}\BibitemShut {NoStop}%
\bibitem [{\citenamefont {Schmitt}\ \emph {et~al.}(1993)\citenamefont
  {Schmitt}, \citenamefont {Grunewald}, \citenamefont {Decher}, \citenamefont
  {Pershan}, \citenamefont {Kjaer},\ and\ \citenamefont {Losche}}]{Schmitt93}%
  \BibitemOpen
  \bibfield  {author} {\bibinfo {author} {\bibfnamefont {J.}~\bibnamefont
  {Schmitt}}, \bibinfo {author} {\bibfnamefont {T.}~\bibnamefont {Grunewald}},
  \bibinfo {author} {\bibfnamefont {G.}~\bibnamefont {Decher}}, \bibinfo
  {author} {\bibfnamefont {P.}~\bibnamefont {Pershan}}, \bibinfo {author}
  {\bibfnamefont {K.}~\bibnamefont {Kjaer}}, \ and\ \bibinfo {author}
  {\bibfnamefont {M.}~\bibnamefont {Losche}},\ }\href@noop {} {\bibfield
  {journal} {\bibinfo  {journal} {Macromolecules}\ }\textbf {\bibinfo {volume}
  {26}},\ \bibinfo {pages} {7058} (\bibinfo {year} {1993})}\BibitemShut
  {NoStop}%
\bibitem [{\citenamefont {Decher}(1997)}]{Decher97}%
  \BibitemOpen
  \bibfield  {author} {\bibinfo {author} {\bibfnamefont {G.}~\bibnamefont
  {Decher}},\ }\href@noop {} {\bibfield  {journal} {\bibinfo  {journal}
  {Science}\ }\textbf {\bibinfo {volume} {277}},\ \bibinfo {pages} {1232}
  (\bibinfo {year} {1997})}\BibitemShut {NoStop}%
\bibitem [{\citenamefont {Rycenga}, \citenamefont {Camargo},\ and\
  \citenamefont {Xia}(2009)}]{Rycenga09}%
  \BibitemOpen
  \bibfield  {author} {\bibinfo {author} {\bibfnamefont {M.}~\bibnamefont
  {Rycenga}}, \bibinfo {author} {\bibfnamefont {P.~H.~C.}\ \bibnamefont
  {Camargo}}, \ and\ \bibinfo {author} {\bibfnamefont {Y.}~\bibnamefont
  {Xia}},\ }\href@noop {} {\bibfield  {journal} {\bibinfo  {journal} {Soft
  Matter}\ }\textbf {\bibinfo {volume} {5}},\ \bibinfo {pages} {1129} (\bibinfo
  {year} {2009})}\BibitemShut {NoStop}%
\bibitem [{\citenamefont {Qi}\ \emph {et~al.}(2012)\citenamefont {Qi},
  \citenamefont {de~Graaf}, \citenamefont {Qiao}, \citenamefont {Marras},
  \citenamefont {Manna},\ and\ \citenamefont {Dijkstra}}]{Qi12}%
  \BibitemOpen
  \bibfield  {author} {\bibinfo {author} {\bibfnamefont {W.}~\bibnamefont
  {Qi}}, \bibinfo {author} {\bibfnamefont {J.}~\bibnamefont {de~Graaf}},
  \bibinfo {author} {\bibfnamefont {F.}~\bibnamefont {Qiao}}, \bibinfo {author}
  {\bibfnamefont {S.}~\bibnamefont {Marras}}, \bibinfo {author} {\bibfnamefont
  {L.}~\bibnamefont {Manna}}, \ and\ \bibinfo {author} {\bibfnamefont
  {M.}~\bibnamefont {Dijkstra}},\ }\href@noop {} {\bibfield  {journal}
  {\bibinfo  {journal} {Nano Letters}\ }\textbf {\bibinfo {volume} {12}},\
  \bibinfo {pages} {5299} (\bibinfo {year} {2012})}\BibitemShut {NoStop}%
\bibitem [{\citenamefont {Frenkel}\ and\ \citenamefont
  {Eppenga}(1985)}]{Frenkel-Eppenga}%
  \BibitemOpen
  \bibfield  {author} {\bibinfo {author} {\bibfnamefont {D.}~\bibnamefont
  {Frenkel}}\ and\ \bibinfo {author} {\bibfnamefont {R.}~\bibnamefont
  {Eppenga}},\ }\href@noop {} {\bibfield  {journal} {\bibinfo  {journal} {Phys.
  Rev. A}\ }\textbf {\bibinfo {volume} {31}},\ \bibinfo {pages} {1776}
  (\bibinfo {year} {1985})}\BibitemShut {NoStop}%
\bibitem [{\citenamefont {Cuesta}\ and\ \citenamefont
  {Frenkel}(1990)}]{Cuesta90}%
  \BibitemOpen
  \bibfield  {author} {\bibinfo {author} {\bibfnamefont {J.~A.}\ \bibnamefont
  {Cuesta}}\ and\ \bibinfo {author} {\bibfnamefont {D.}~\bibnamefont
  {Frenkel}},\ }\href@noop {} {\bibfield  {journal} {\bibinfo  {journal} {Phys.
  Rev. A}\ }\textbf {\bibinfo {volume} {42}},\ \bibinfo {pages} {2126}
  (\bibinfo {year} {1990})}\BibitemShut {NoStop}%
\bibitem [{\citenamefont {Donev}\ \emph {et~al.}(2006)\citenamefont {Donev},
  \citenamefont {Burton}, \citenamefont {Stillinger},\ and\ \citenamefont
  {Torquato}}]{Donev06}%
  \BibitemOpen
  \bibfield  {author} {\bibinfo {author} {\bibfnamefont {A.}~\bibnamefont
  {Donev}}, \bibinfo {author} {\bibfnamefont {J.}~\bibnamefont {Burton}},
  \bibinfo {author} {\bibfnamefont {F.~H.}\ \bibnamefont {Stillinger}}, \ and\
  \bibinfo {author} {\bibfnamefont {S.}~\bibnamefont {Torquato}},\ }\href@noop
  {} {\bibfield  {journal} {\bibinfo  {journal} {Phys. Rev. B.}\ }\textbf
  {\bibinfo {volume} {73}},\ \bibinfo {pages} {054109} (\bibinfo {year}
  {2006})}\BibitemShut {NoStop}%
\bibitem [{\citenamefont {Avendano}\ and\ \citenamefont
  {Escobedo}(2012)}]{Avendano12}%
  \BibitemOpen
  \bibfield  {author} {\bibinfo {author} {\bibfnamefont {C.}~\bibnamefont
  {Avendano}}\ and\ \bibinfo {author} {\bibfnamefont {F.~A.}\ \bibnamefont
  {Escobedo}},\ }\href@noop {} {\bibfield  {journal} {\bibinfo  {journal} {Soft
  Matter}\ }\textbf {\bibinfo {volume} {8}},\ \bibinfo {pages} {4675} (\bibinfo
  {year} {2012})}\BibitemShut {NoStop}%
\bibitem [{\citenamefont {Shah}\ \emph {et~al.}(2012)\citenamefont {Shah},
  \citenamefont {Kang}, \citenamefont {Kohlstedt}, \citenamefont {Ahn},
  \citenamefont {Glotzer}, \citenamefont {Monroe},\ and\ \citenamefont
  {Solomon}}]{Shah12}%
  \BibitemOpen
  \bibfield  {author} {\bibinfo {author} {\bibfnamefont {A.~A.}\ \bibnamefont
  {Shah}}, \bibinfo {author} {\bibfnamefont {H.}~\bibnamefont {Kang}}, \bibinfo
  {author} {\bibfnamefont {K.~L.}\ \bibnamefont {Kohlstedt}}, \bibinfo {author}
  {\bibfnamefont {K.~H.}\ \bibnamefont {Ahn}}, \bibinfo {author} {\bibfnamefont
  {S.~C.}\ \bibnamefont {Glotzer}}, \bibinfo {author} {\bibfnamefont {C.~W.}\
  \bibnamefont {Monroe}}, \ and\ \bibinfo {author} {\bibfnamefont {M.~J.}\
  \bibnamefont {Solomon}},\ }\href@noop {} {\bibfield  {journal} {\bibinfo
  {journal} {Small}\ }\textbf {\bibinfo {volume} {8}},\ \bibinfo {pages} {1551}
  (\bibinfo {year} {2012})}\BibitemShut {NoStop}%
\bibitem [{\citenamefont {Qi}\ \emph {et~al.}(2013)\citenamefont {Qi},
  \citenamefont {de~Graaf}, \citenamefont {Qiao}, \citenamefont {Marras},
  \citenamefont {Manna},\ and\ \citenamefont {Dijkstra}}]{Qi13}%
  \BibitemOpen
  \bibfield  {author} {\bibinfo {author} {\bibfnamefont {W.}~\bibnamefont
  {Qi}}, \bibinfo {author} {\bibfnamefont {J.}~\bibnamefont {de~Graaf}},
  \bibinfo {author} {\bibfnamefont {F.}~\bibnamefont {Qiao}}, \bibinfo {author}
  {\bibfnamefont {S.}~\bibnamefont {Marras}}, \bibinfo {author} {\bibfnamefont
  {L.}~\bibnamefont {Manna}}, \ and\ \bibinfo {author} {\bibfnamefont
  {M.}~\bibnamefont {Dijkstra}},\ }\href@noop {} {\bibfield  {journal}
  {\bibinfo  {journal} {J. Chem. Phys.}\ }\textbf {\bibinfo {volume} {138}},\
  \bibinfo {pages} {154504} (\bibinfo {year} {2013})}\BibitemShut {NoStop}%
\bibitem [{\citenamefont {Damasceno}, \citenamefont {Engel},\ and\
  \citenamefont {Glotzer}(2012)}]{Damasceno12}%
  \BibitemOpen
  \bibfield  {author} {\bibinfo {author} {\bibfnamefont {P.~F.}\ \bibnamefont
  {Damasceno}}, \bibinfo {author} {\bibfnamefont {M.}~\bibnamefont {Engel}}, \
  and\ \bibinfo {author} {\bibfnamefont {S.~C.}\ \bibnamefont {Glotzer}},\
  }\href@noop {} {\bibfield  {journal} {\bibinfo  {journal} {ACS Nano}\
  }\textbf {\bibinfo {volume} {6}},\ \bibinfo {pages} {609} (\bibinfo {year}
  {2012})}\BibitemShut {NoStop}%
\bibitem [{\citenamefont {van Anders}\ \emph {et~al.}(2014)\citenamefont {van
  Anders}, \citenamefont {Ahmed}, \citenamefont {Smith}, \citenamefont
  {Engel},\ and\ \citenamefont {Glotzer}}]{van-Anders14}%
  \BibitemOpen
  \bibfield  {author} {\bibinfo {author} {\bibfnamefont {G.}~\bibnamefont {van
  Anders}}, \bibinfo {author} {\bibfnamefont {N.~K.}\ \bibnamefont {Ahmed}},
  \bibinfo {author} {\bibfnamefont {R.}~\bibnamefont {Smith}}, \bibinfo
  {author} {\bibfnamefont {M.}~\bibnamefont {Engel}}, \ and\ \bibinfo {author}
  {\bibfnamefont {S.~C.}\ \bibnamefont {Glotzer}},\ }\href@noop {} {\bibfield
  {journal} {\bibinfo  {journal} {ACS Nano}\ ,\ \bibinfo {pages}
  {DOI:10.1021/nn4057353}} (\bibinfo {year} {2014})}\BibitemShut {NoStop}%
\bibitem [{\citenamefont {Kosterlitz}\ and\ \citenamefont
  {Thouless}(1973)}]{Kosterlitz-Thouless}%
  \BibitemOpen
  \bibfield  {author} {\bibinfo {author} {\bibfnamefont {J.~M.}\ \bibnamefont
  {Kosterlitz}}\ and\ \bibinfo {author} {\bibfnamefont {D.~J.}\ \bibnamefont
  {Thouless}},\ }\href@noop {} {\bibfield  {journal} {\bibinfo  {journal} {J.
  Phys. C}\ }\textbf {\bibinfo {volume} {6}},\ \bibinfo {pages} {1181}
  (\bibinfo {year} {1973})}\BibitemShut {NoStop}%
\bibitem [{\citenamefont {Halperin}\ and\ \citenamefont
  {Nelson}(1978)}]{Halperin-Nelson}%
  \BibitemOpen
  \bibfield  {author} {\bibinfo {author} {\bibfnamefont {B.~I.}\ \bibnamefont
  {Halperin}}\ and\ \bibinfo {author} {\bibfnamefont {D.~R.}\ \bibnamefont
  {Nelson}},\ }\href@noop {} {\bibfield  {journal} {\bibinfo  {journal} {Phys.
  Rev. Lett.}\ }\textbf {\bibinfo {volume} {41}},\ \bibinfo {pages} {121}
  (\bibinfo {year} {1978})}\BibitemShut {NoStop}%
\bibitem [{\citenamefont {Straley}(1971)}]{Straley71}%
  \BibitemOpen
  \bibfield  {author} {\bibinfo {author} {\bibfnamefont {J.~P.}\ \bibnamefont
  {Straley}},\ }\href@noop {} {\bibfield  {journal} {\bibinfo  {journal} {Phys.
  Rev. A}\ }\textbf {\bibinfo {volume} {4}},\ \bibinfo {pages} {675} (\bibinfo
  {year} {1971})}\BibitemShut {NoStop}%
\bibitem [{\citenamefont {Bates}\ and\ \citenamefont
  {Frenkel}(2000)}]{Bates-Frenkel}%
  \BibitemOpen
  \bibfield  {author} {\bibinfo {author} {\bibfnamefont {M.~A.}\ \bibnamefont
  {Bates}}\ and\ \bibinfo {author} {\bibfnamefont {D.}~\bibnamefont
  {Frenkel}},\ }\href@noop {} {\bibfield  {journal} {\bibinfo  {journal} {J.
  Chem. Phys.}\ }\textbf {\bibinfo {volume} {112}},\ \bibinfo {pages} {10034}
  (\bibinfo {year} {2000})}\BibitemShut {NoStop}%
\bibitem [{\citenamefont {Zheng}\ and\ \citenamefont {Han}(2010)}]{Zheng10}%
  \BibitemOpen
  \bibfield  {author} {\bibinfo {author} {\bibfnamefont {Z.}~\bibnamefont
  {Zheng}}\ and\ \bibinfo {author} {\bibfnamefont {Y.}~\bibnamefont {Han}},\
  }\href@noop {} {\bibfield  {journal} {\bibinfo  {journal} {J. Chem. Phys.}\
  }\textbf {\bibinfo {volume} {133}},\ \bibinfo {pages} {124509} (\bibinfo
  {year} {2010})}\BibitemShut {NoStop}%
\bibitem [{\citenamefont {Xu}\ \emph {et~al.}(2013)\citenamefont {Xu},
  \citenamefont {Li}, \citenamefont {Sun},\ and\ \citenamefont {An}}]{Xu13}%
  \BibitemOpen
  \bibfield  {author} {\bibinfo {author} {\bibfnamefont {W.~S.}\ \bibnamefont
  {Xu}}, \bibinfo {author} {\bibfnamefont {Y.~W.}\ \bibnamefont {Li}}, \bibinfo
  {author} {\bibfnamefont {Z.~Y.}\ \bibnamefont {Sun}}, \ and\ \bibinfo
  {author} {\bibfnamefont {L.~J.}\ \bibnamefont {An}},\ }\href@noop {}
  {\bibfield  {journal} {\bibinfo  {journal} {J. Chem. Phys.}\ }\textbf
  {\bibinfo {volume} {139}},\ \bibinfo {pages} {024501} (\bibinfo {year}
  {2013})}\BibitemShut {NoStop}%
\bibitem [{\citenamefont {Jaster}(1999)}]{Jaster99}%
  \BibitemOpen
  \bibfield  {author} {\bibinfo {author} {\bibfnamefont {A.}~\bibnamefont
  {Jaster}},\ }\href@noop {} {\bibfield  {journal} {\bibinfo  {journal} {Phys.
  Rev. E.}\ }\textbf {\bibinfo {volume} {59}},\ \bibinfo {pages} {2594}
  (\bibinfo {year} {1999})}\BibitemShut {NoStop}%
\bibitem [{\citenamefont {Bernard}\ and\ \citenamefont
  {Krauth}(2011)}]{Bernard11}%
  \BibitemOpen
  \bibfield  {author} {\bibinfo {author} {\bibfnamefont {E.~P.}\ \bibnamefont
  {Bernard}}\ and\ \bibinfo {author} {\bibfnamefont {W.}~\bibnamefont
  {Krauth}},\ }\href@noop {} {\bibfield  {journal} {\bibinfo  {journal} {Phys.
  Rev. Lett.}\ }\textbf {\bibinfo {volume} {107}},\ \bibinfo {pages} {155704}
  (\bibinfo {year} {2011})}\BibitemShut {NoStop}%
\bibitem [{\citenamefont {Vieillard-Baron}(1972)}]{Vieillard-Baron72}%
  \BibitemOpen
  \bibfield  {author} {\bibinfo {author} {\bibfnamefont {J.}~\bibnamefont
  {Vieillard-Baron}},\ }\href@noop {} {\bibfield  {journal} {\bibinfo
  {journal} {J. Chem. Phys.}\ }\textbf {\bibinfo {volume} {56}},\ \bibinfo
  {pages} {4729} (\bibinfo {year} {1972})}\BibitemShut {NoStop}%
\bibitem [{\citenamefont {Levesque}, \citenamefont {Schiff},\ and\
  \citenamefont {Vieillard-Baron}(1969)}]{Levesque69}%
  \BibitemOpen
  \bibfield  {author} {\bibinfo {author} {\bibfnamefont {D.}~\bibnamefont
  {Levesque}}, \bibinfo {author} {\bibfnamefont {D.}~\bibnamefont {Schiff}}, \
  and\ \bibinfo {author} {\bibfnamefont {J.}~\bibnamefont {Vieillard-Baron}},\
  }\href@noop {} {\bibfield  {journal} {\bibinfo  {journal} {J. Chem. Phys.}\
  }\textbf {\bibinfo {volume} {51}},\ \bibinfo {pages} {3625} (\bibinfo {year}
  {1969})}\BibitemShut {NoStop}%
\bibitem [{\citenamefont {Foulaadvand}\ and\ \citenamefont
  {Yarifard}(2013)}]{Foulaadvand13}%
  \BibitemOpen
  \bibfield  {author} {\bibinfo {author} {\bibfnamefont {M.~E.}\ \bibnamefont
  {Foulaadvand}}\ and\ \bibinfo {author} {\bibfnamefont {M.}~\bibnamefont
  {Yarifard}},\ }\href@noop {} {\bibfield  {journal} {\bibinfo  {journal}
  {Phys. Rev. E.}\ }\textbf {\bibinfo {volume} {88}},\ \bibinfo {pages}
  {052504} (\bibinfo {year} {2013})}\BibitemShut {NoStop}%
\bibitem [{\citenamefont {Rickayzen}(1998)}]{Rickayzen98}%
  \BibitemOpen
  \bibfield  {author} {\bibinfo {author} {\bibfnamefont {G.}~\bibnamefont
  {Rickayzen}},\ }\href@noop {} {\bibfield  {journal} {\bibinfo  {journal}
  {Mol. Phys.}\ }\textbf {\bibinfo {volume} {95}},\ \bibinfo {pages} {393}
  (\bibinfo {year} {1998})}\BibitemShut {NoStop}%
\bibitem [{\citenamefont {Perram}\ \emph {et~al.}(1984)\citenamefont {Perram},
  \citenamefont {Wertheim}, \citenamefont {Lebowitz},\ and\ \citenamefont
  {Williams}}]{Perram84}%
  \BibitemOpen
  \bibfield  {author} {\bibinfo {author} {\bibfnamefont {J.~W.}\ \bibnamefont
  {Perram}}, \bibinfo {author} {\bibfnamefont {M.~S.}\ \bibnamefont
  {Wertheim}}, \bibinfo {author} {\bibfnamefont {J.~L.}\ \bibnamefont
  {Lebowitz}}, \ and\ \bibinfo {author} {\bibfnamefont {G.~O.}\ \bibnamefont
  {Williams}},\ }\href@noop {} {\bibfield  {journal} {\bibinfo  {journal}
  {Chem. Phys. Lett.}\ }\textbf {\bibinfo {volume} {105}},\ \bibinfo {pages}
  {277} (\bibinfo {year} {1984})}\BibitemShut {NoStop}%
\bibitem [{\citenamefont {Perram}\ and\ \citenamefont
  {Wertheim}(1985)}]{Perram85}%
  \BibitemOpen
  \bibfield  {author} {\bibinfo {author} {\bibfnamefont {J.~W.}\ \bibnamefont
  {Perram}}\ and\ \bibinfo {author} {\bibfnamefont {M.~S.}\ \bibnamefont
  {Wertheim}},\ }\href@noop {} {\bibfield  {journal} {\bibinfo  {journal} {J.
  Comput. Phys.}\ }\textbf {\bibinfo {volume} {58}},\ \bibinfo {pages} {409}
  (\bibinfo {year} {1985})}\BibitemShut {NoStop}%
\bibitem [{\citenamefont {de~J.~Guevara-Rodr\'{i}guez}\ and\ \citenamefont
  {Odriozola}(2011)}]{GuevaraHE}%
  \BibitemOpen
  \bibfield  {author} {\bibinfo {author} {\bibfnamefont {F.}~\bibnamefont
  {de~J.~Guevara-Rodr\'{i}guez}}\ and\ \bibinfo {author} {\bibfnamefont
  {G.}~\bibnamefont {Odriozola}},\ }\href@noop {} {\bibfield  {journal}
  {\bibinfo  {journal} {J. Chem. Phys.}\ }\textbf {\bibinfo {volume} {135}},\
  \bibinfo {pages} {084508} (\bibinfo {year} {2011})}\BibitemShut {NoStop}%
\bibitem [{\citenamefont {Bautista-Carbajal}, \citenamefont {Moncho-Jord\'a},\
  and\ \citenamefont {Odriozola}(2013)}]{Bautista13}%
  \BibitemOpen
  \bibfield  {author} {\bibinfo {author} {\bibfnamefont {G.}~\bibnamefont
  {Bautista-Carbajal}}, \bibinfo {author} {\bibfnamefont {A.}~\bibnamefont
  {Moncho-Jord\'a}}, \ and\ \bibinfo {author} {\bibfnamefont {G.}~\bibnamefont
  {Odriozola}},\ }\href@noop {} {\bibfield  {journal} {\bibinfo  {journal} {J.
  Chem. Phys.}\ }\textbf {\bibinfo {volume} {138}},\ \bibinfo {pages} {064501}
  (\bibinfo {year} {2013})}\BibitemShut {NoStop}%
\bibitem [{\citenamefont {Marinari}\ and\ \citenamefont
  {Parisi}(1992)}]{Marinari92}%
  \BibitemOpen
  \bibfield  {author} {\bibinfo {author} {\bibfnamefont {E.}~\bibnamefont
  {Marinari}}\ and\ \bibinfo {author} {\bibfnamefont {G.}~\bibnamefont
  {Parisi}},\ }\href@noop {} {\bibfield  {journal} {\bibinfo  {journal}
  {Europhys. Lett.}\ }\textbf {\bibinfo {volume} {19}},\ \bibinfo {pages} {451}
  (\bibinfo {year} {1992})}\BibitemShut {NoStop}%
\bibitem [{\citenamefont {Lyubartsev}\ \emph {et~al.}(1992)\citenamefont
  {Lyubartsev}, \citenamefont {Martinovski}, \citenamefont {Shevkunov},\ and\
  \citenamefont {Vorontsov-Velyaminov}}]{Lyubartsev92}%
  \BibitemOpen
  \bibfield  {author} {\bibinfo {author} {\bibfnamefont {A.~P.}\ \bibnamefont
  {Lyubartsev}}, \bibinfo {author} {\bibfnamefont {A.~A.}\ \bibnamefont
  {Martinovski}}, \bibinfo {author} {\bibfnamefont {S.~V.}\ \bibnamefont
  {Shevkunov}}, \ and\ \bibinfo {author} {\bibfnamefont {P.~N.}\ \bibnamefont
  {Vorontsov-Velyaminov}},\ }\href@noop {} {\bibfield  {journal} {\bibinfo
  {journal} {J. Chem. Phys.}\ }\textbf {\bibinfo {volume} {96}},\ \bibinfo
  {pages} {1776} (\bibinfo {year} {1992})}\BibitemShut {NoStop}%
\bibitem [{\citenamefont {Hukushima}\ and\ \citenamefont
  {Nemoto}(1996)}]{hukushima96}%
  \BibitemOpen
  \bibfield  {author} {\bibinfo {author} {\bibfnamefont {K.}~\bibnamefont
  {Hukushima}}\ and\ \bibinfo {author} {\bibfnamefont {K.}~\bibnamefont
  {Nemoto}},\ }\href@noop {} {\bibfield  {journal} {\bibinfo  {journal} {J.
  Phys. Soc. Jpn.}\ }\textbf {\bibinfo {volume} {65}},\ \bibinfo {pages} {1604}
  (\bibinfo {year} {1996})}\BibitemShut {NoStop}%
\bibitem [{\citenamefont {Odriozola}(2009)}]{Odriozola09}%
  \BibitemOpen
  \bibfield  {author} {\bibinfo {author} {\bibfnamefont {G.}~\bibnamefont
  {Odriozola}},\ }\href@noop {} {\bibfield  {journal} {\bibinfo  {journal} {J.
  Chem. Phys.}\ }\textbf {\bibinfo {volume} {131}},\ \bibinfo {pages} {144107}
  (\bibinfo {year} {2009})}\BibitemShut {NoStop}%
\bibitem [{\citenamefont {Okabe}\ \emph {et~al.}(2001)\citenamefont {Okabe},
  \citenamefont {Kawata}, \citenamefont {Okamoto},\ and\ \citenamefont
  {Mikami}}]{Okabe01}%
  \BibitemOpen
  \bibfield  {author} {\bibinfo {author} {\bibfnamefont {T.}~\bibnamefont
  {Okabe}}, \bibinfo {author} {\bibfnamefont {M.}~\bibnamefont {Kawata}},
  \bibinfo {author} {\bibfnamefont {Y.}~\bibnamefont {Okamoto}}, \ and\
  \bibinfo {author} {\bibfnamefont {M.}~\bibnamefont {Mikami}},\ }\href@noop {}
  {\bibfield  {journal} {\bibinfo  {journal} {Chem. Phys. Lett.}\ }\textbf
  {\bibinfo {volume} {335}},\ \bibinfo {pages} {435} (\bibinfo {year}
  {2001})}\BibitemShut {NoStop}%
\bibitem [{\citenamefont {Donev}\ \emph {et~al.}(2004)\citenamefont {Donev},
  \citenamefont {Stillinger}, \citenamefont {Chaikin},\ and\ \citenamefont
  {Torquato}}]{Donev04a}%
  \BibitemOpen
  \bibfield  {author} {\bibinfo {author} {\bibfnamefont {A.}~\bibnamefont
  {Donev}}, \bibinfo {author} {\bibfnamefont {F.~H.}\ \bibnamefont
  {Stillinger}}, \bibinfo {author} {\bibfnamefont {P.~M.}\ \bibnamefont
  {Chaikin}}, \ and\ \bibinfo {author} {\bibfnamefont {S.}~\bibnamefont
  {Torquato}},\ }\href@noop {} {\bibfield  {journal} {\bibinfo  {journal}
  {Phys. Rev. Lett.}\ }\textbf {\bibinfo {volume} {92}},\ \bibinfo {pages}
  {255506} (\bibinfo {year} {2004})}\BibitemShut {NoStop}%
\bibitem [{\citenamefont {Rathore}, \citenamefont {Chopra},\ and\ \citenamefont
  {de~Pablo}(2005)}]{Rathore05}%
  \BibitemOpen
  \bibfield  {author} {\bibinfo {author} {\bibfnamefont {N.}~\bibnamefont
  {Rathore}}, \bibinfo {author} {\bibfnamefont {M.}~\bibnamefont {Chopra}}, \
  and\ \bibinfo {author} {\bibfnamefont {J.~J.}\ \bibnamefont {de~Pablo}},\
  }\href@noop {} {\bibfield  {journal} {\bibinfo  {journal} {J. Chem. Phys.}\
  }\textbf {\bibinfo {volume} {122}},\ \bibinfo {pages} {024111} (\bibinfo
  {year} {2005})}\BibitemShut {NoStop}%
\bibitem [{\citenamefont {Donev}, \citenamefont {Torquato},\ and\ \citenamefont
  {Stillinger}(2005)}]{Donev05a}%
  \BibitemOpen
  \bibfield  {author} {\bibinfo {author} {\bibfnamefont {A.}~\bibnamefont
  {Donev}}, \bibinfo {author} {\bibfnamefont {S.}~\bibnamefont {Torquato}}, \
  and\ \bibinfo {author} {\bibfnamefont {F.~H.}\ \bibnamefont {Stillinger}},\
  }\href@noop {} {\bibfield  {journal} {\bibinfo  {journal} {J. Comput. Phys.}\
  }\textbf {\bibinfo {volume} {202}},\ \bibinfo {pages} {737} (\bibinfo {year}
  {2005})}\BibitemShut {NoStop}%
\bibitem [{\citenamefont {Ferrenberg}\ and\ \citenamefont
  {Swendsen}(1988)}]{Ferrenberg88}%
  \BibitemOpen
  \bibfield  {author} {\bibinfo {author} {\bibfnamefont {A.~M.}\ \bibnamefont
  {Ferrenberg}}\ and\ \bibinfo {author} {\bibfnamefont {R.~H.}\ \bibnamefont
  {Swendsen}},\ }\href@noop {} {\bibfield  {journal} {\bibinfo  {journal}
  {Phys. Rev. Lett.}\ }\textbf {\bibinfo {volume} {61}},\ \bibinfo {pages}
  {2635} (\bibinfo {year} {1988})}\BibitemShut {NoStop}%
\bibitem [{\citenamefont {Ferrenberg}\ and\ \citenamefont
  {Swendsen}(1989)}]{Ferrenberg89}%
  \BibitemOpen
  \bibfield  {author} {\bibinfo {author} {\bibfnamefont {A.~M.}\ \bibnamefont
  {Ferrenberg}}\ and\ \bibinfo {author} {\bibfnamefont {R.~H.}\ \bibnamefont
  {Swendsen}},\ }\href@noop {} {\bibfield  {journal} {\bibinfo  {journal}
  {Phys. Rev. Lett.}\ }\textbf {\bibinfo {volume} {63}},\ \bibinfo {pages}
  {1195} (\bibinfo {year} {1989})}\BibitemShut {NoStop}%
\bibitem [{\citenamefont {Varga}\ and\ \citenamefont {Szalai}(1998)}]{Varga98}%
  \BibitemOpen
  \bibfield  {author} {\bibinfo {author} {\bibfnamefont {S.}~\bibnamefont
  {Varga}}\ and\ \bibinfo {author} {\bibfnamefont {I.}~\bibnamefont {Szalai}},\
  }\href@noop {} {\bibfield  {journal} {\bibinfo  {journal} {Mol. Phys.}\
  }\textbf {\bibinfo {volume} {95}},\ \bibinfo {pages} {515} (\bibinfo {year}
  {1998})}\BibitemShut {NoStop}%
\bibitem [{\citenamefont {Boubl\'{i}k}(2011)}]{Boublik11}%
  \BibitemOpen
  \bibfield  {author} {\bibinfo {author} {\bibfnamefont {T.}~\bibnamefont
  {Boubl\'{i}k}},\ }\href@noop {} {\bibfield  {journal} {\bibinfo  {journal}
  {Mol. Phys.}\ }\textbf {\bibinfo {volume} {109}},\ \bibinfo {pages} {1575}
  (\bibinfo {year} {2011})}\BibitemShut {NoStop}%
\bibitem [{\citenamefont {Bolhuis}\ and\ \citenamefont
  {Frenkel}(1997)}]{Bolhuis97}%
  \BibitemOpen
  \bibfield  {author} {\bibinfo {author} {\bibfnamefont {P.}~\bibnamefont
  {Bolhuis}}\ and\ \bibinfo {author} {\bibfnamefont {D.}~\bibnamefont
  {Frenkel}},\ }\href@noop {} {\bibfield  {journal} {\bibinfo  {journal} {J.
  Chem. Phys.}\ }\textbf {\bibinfo {volume} {106}},\ \bibinfo {pages} {666}
  (\bibinfo {year} {1997})}\BibitemShut {NoStop}%
\bibitem [{\citenamefont {Wensink}\ and\ \citenamefont
  {Lekkerkerker}(2009)}]{Wensink09}%
  \BibitemOpen
  \bibfield  {author} {\bibinfo {author} {\bibfnamefont {H.~H.}\ \bibnamefont
  {Wensink}}\ and\ \bibinfo {author} {\bibfnamefont {H.~N.~W.}\ \bibnamefont
  {Lekkerkerker}},\ }\href@noop {} {\bibfield  {journal} {\bibinfo  {journal}
  {Mol. Phys.}\ }\textbf {\bibinfo {volume} {107}},\ \bibinfo {pages} {2111}
  (\bibinfo {year} {2009})}\BibitemShut {NoStop}%
\bibitem [{\citenamefont {Marechal}\ \emph {et~al.}(2011)\citenamefont
  {Marechal}, \citenamefont {Cuetos}, \citenamefont {Mart\'inez-Haya},\ and\
  \citenamefont {Dijkstra}}]{Marechal11}%
  \BibitemOpen
  \bibfield  {author} {\bibinfo {author} {\bibfnamefont {M.}~\bibnamefont
  {Marechal}}, \bibinfo {author} {\bibfnamefont {A.}~\bibnamefont {Cuetos}},
  \bibinfo {author} {\bibfnamefont {B.}~\bibnamefont {Mart\'inez-Haya}}, \ and\
  \bibinfo {author} {\bibfnamefont {M.}~\bibnamefont {Dijkstra}},\ }\href@noop
  {} {\bibfield  {journal} {\bibinfo  {journal} {J. Chem. Phys.}\ }\textbf
  {\bibinfo {volume} {134}},\ \bibinfo {pages} {094501} (\bibinfo {year}
  {2011})}\BibitemShut {NoStop} %
\end{thebibliography}

%

\end{document}